\documentclass[letterpaper,12pt]{article}
\pdfoutput=1

\usepackage{jheppub}
\usepackage{amsmath}
\usepackage{graphicx}
\usepackage{subfig}
\usepackage{cancel}
\usepackage[dvipsnames]{xcolor}
\usepackage{color}
\usepackage{mathrsfs}

\usepackage{amssymb}
\usepackage{epstopdf}
\def\({\left(}
\def\){\right)}
\def\[{\left[}
\def\]{\right]}

\newpage

\title{Violating the Quantum Focusing Conjecture and Quantum Covariant Entropy Bound in $d\ge 5$ dimensions}

\author[a]{Zicao Fu}
\author[b,c]{Jason Koeller}
\author[a]{Donald Marolf}

\affiliation[a]{Department of Physics, University of California, Santa Barbara, CA 93106, USA}
\affiliation[b]{Center for Theoretical Physics and Department of Physics,\\
University of California, Berkeley, CA 94720, USA}
\affiliation[c]{Lawrence Berkeley National Laboratory, Berkeley, CA 94720, USA}

\emailAdd{zicaofu@physics.ucsb.edu}
\emailAdd{jkoeller@berkeley.edu}
\emailAdd{marolf@physics.ucsb.edu}

\abstract{We study the Quantum Focussing Conjecture (QFC) in curved spacetime. Noting that quantum corrections from integrating out massive fields generally induce a Gauss-Bonnet term, we study Einstein-Hilbert-Gauss-Bonnet gravity and show for $d\ge 5$ spacetime dimensions that weakly-curved solutions can violate the associated QFC for either sign of the Gauss-Bonnet coupling. The nature of the violation shows that -- so long as the Gauss-Bonnet coupling is non-zero -- it will continue to arise for local effective actions containing arbitrary further higher curvature terms, and when gravity is coupled to generic $d\ge 5$ theories of massive quantum fields. The argument also implies violations of a recently-conjectured form of the generalized covariant entropy bound. The possible validity of the QFC and covariant entropy bound in $d\le 4$ spacetime dimensions remains open.}

\begin{document}
\maketitle

\section{Introduction}
\label{sec:Intro}

The gravitational focussing theorem plays a key role in the modern understanding of General Relativity.  This key results states (see e.g. \cite{Wald:1984rg}) that the expansion of null congruences cannot increase toward the future in any solution to Einstein-Hilbert gravity sourced by matter satisfying the Null Energy Condition (NEC).  It leads to the second law of black hole thermodynamics \cite{Hawking:1971tu},  singularity theorems \cite{Penrose:1964wq,Hawking:1973uf}, the chronology protection theorem \cite{Hawking:1991nk}, topological censorship \cite{Friedman:1993ty}, and other fundamental results.   It also guarantees essential properties of holographic entanglement entropy \cite{Wall:2012uf,Headrick:2014cta} in the context of gauge/gravity duality.

However, the null energy condition is known to be violated by quantum effects \cite{Epstein:1965zza}.  This then raises the question of whether quantum corrections might enable fundamentally new and perhaps pathological gravitational phenomena.  Indeed, it was recently established that traversable wormholes can be constructed in this way \cite{Gao:2016bin}.  On the other hand, the conjectured Generalized Second Law of thermodynamics (GSL) would both limit the utility of traversable wormholes and prohibit even more troubling exotic physics \cite{Wall:2013uza}.

Motivated in part by the GSL, and also in part by the covariant entropy conjecture \cite{Bousso:1999xy}, it was suggested in \cite{Bousso:2015mna} that a generalization of the focussing theorem might continue to hold at the quantum level.   Known as the Quantum Focussing Conjecture (QFC), it would imply both the GSL (for any causal horizon) and a form \cite{Bousso:2015mna} of the covariant entropy bound of \cite{Bousso:1999xy} related to the version discussed by Strominger and Thompson \cite{Strominger:2003br}.

The QFC is formulated by noting that the expansion $\theta$ of any null congruence can be expressed as a first functional derivative of the area of cuts of the congruence, and that Einstein-Hilbert gravity associates a Bekenstein-Hawking entropy $S_{\rm BH} = A/4G
$ with many surfaces of area $A$.    In particular, given a region ${\cal R}$ with boundary $\Sigma = \partial {\cal R}$ in some Cauchy surface, and also given a null congruence $N$ orthogonal to $\Sigma$, we have
\begin{align}
\label{exp}
	\theta[\Sigma,y] = \frac{4G}{\sqrt{\tilde{h}}} \frac{\delta S_{\rm BH}}{\delta \Sigma(y)},
\end{align}
where $y$ labels the space of null generators, $\delta \Sigma(y)$ is an infinitesimal displacement of the surface along the null generator $y$, and $\tilde{h}$ denotes the determinant of the transverse metric in the $y$-coordinate system on the null congruence $N$.  For semi-classical gravity (and in particular where the metric itself may be treated classically),
ref. \cite{Bousso:2015mna} then defines the generalized expansion $\Theta[\sigma, y]$ by replacing $S_{\rm BH} = A/4G$ in \eqref{exp} with the generalized entropy functional
\begin{equation}\label{SgenDef}
S_\text{gen}=S_\text{grav}+S_\text{out}.
\end{equation}
Here $S_\text{grav}$ is an appropriate gravitational entropy functional (say, from \cite{Dong:2013qoa,Miao:2014nxa,Dong:2015zba,Wall:2015raa}, which coincides with that of \cite{Jacobson:1993xs} for the case studied here) and $S_\text{out}$ is a von Neumann entropy for quantum fields outside the null congruence.\footnote{$S_{\text{out}}$ presumably includes an appropriate set of boundary terms for gauge fields as in e.g. \cite{will_lattice,will14,Ghosh:2015iwa,Aoki:2015bsa,Donnelly:2014fua,Donnelly:2015hxa}.} Finally, the statement of the QFC is simply that $\Theta$ is semi-classically non-decreasing as we push the surface $\Sigma$ toward the future or, in other words, that a corresponding second derivative of $S_{\text{gen}}$ is negative or zero:
\begin{equation}
\label{2fd}
\frac{1}{\sqrt{\tilde h(y)}}\frac{\delta }{\delta \Sigma \left(y_2\right)}\Theta \left[\Sigma; y_1\right]\le 0.
\end{equation}
While \eqref{2fd} is divergent for $y_1=y_2$, and in particular the contribution of the Einstein-Hilbert term to \eqref{2fd} is $\dot{\theta} \delta(y_1 - y_2)$ where $\dot \theta = k^a \nabla_a \theta$, the quantity \eqref{2fd} remains meaningful when treated as a distribution.

As evidence for the QFC, one may recall \cite{Bousso:2015mna} that in Einstein-Hilbert gravity, taking a weakly-gravitating ($G \to 0$) limit implies quantum fields satisfy a so-called Quantum Null Energy Condition (QNEC) generalizing the classical NEC, and that this QNEC has now been established in a variety of contexts \cite{Bousso:2015wca,Koeller:2015qmn}.  In such cases, an associated QFC follows immediately at first order in the coupling $G$ of such theories to Einstein-Hilbert gravity.

However, we argue here that for $d \ge 5$ spacetime dimensions the QFC generally fails.  To do so, we recall that integrating out massive fields typically induces a Gauss-Bonnet term in the gravitational effective action; see e.g. \cite{Shapiro:2008sf}.  Classical Einstein-Hilbert-Gauss-Bonnet gravity is analyzed in section \ref{sec:Violation}, and is shown to violate the QFC at weak curvature for $d \ge 5$.\footnote{Causality violations implying pathologies for non-stringy theories with large Gauss-Bonnet couplings were found in \cite{Camanho:2014apa}. By contrast, we emphasize that the QFC violation found in this paper is present for the less restrictive class of theories containing even a small effective field theory Gauss-Bonnet term.} The form of this violation shows that similar issues arise at the quantum level, and also in the presence of arbitrary higher derivative terms controlled by a single length scale so long as the coefficient of the Gauss-Bonnet term is non-zero.  The QFC is thus violated in generic $d \ge 5$ theories of semi-classical gravity coupled to massive quantum fields, and presumably in the presence of massless quantum fields as well. Our example also leads in section \ref{Bousso} to violations of the generalized covariant entropy bound (also called the quantum Bousso bound) conjectured in \cite{Bousso:2015mna}.\footnote{This conjecture is closely related to the Strominger-Thompson proposal \cite{Strominger:2003br}.}  We close in section \ref{sec:Discussion} with further discussion emphasizing future directions and the possibility that a reformulated QFC and quantum Bousso bound may nevertheless hold.

\section{Violating the QFC in Gauss-Bonnet Gravity }
\label{sec:Violation}

Consider the the Einstein-Hilbert-Gauss-Bonnet action
\begin{equation}
\label{eq:action}
I=\frac{1}{16\pi G}\int d^dx\sqrt{-g}R+\gamma \int d^dx\sqrt{-g}\left(R_{abcd}R^{abcd}-4R_{ab}R^{ab}+R^2\right).
\end{equation}
As noted above, we will first treat this theory classically and identify violations of the associated QFC \eqref{2fd}.  We will then note that explicit quantum corrections are sub-leading in a long-wavelength expansion so our classical violation extends directly to the quantum level.

We work in the weak curvature limit, taking the Weyl tensor to be first order in some small quantity $\epsilon$:
\begin{equation}
\label{eq:weakgravity}
C_{abcd}=O\left(\epsilon \right).
\end{equation}
In this limit,  iteratively solving the equation of motion yields
\begin{equation}
\label{eq:eom}
R_{ab}=\frac{16\pi G\gamma }{d-2}C_{cdef}C^{cdef}g_{ab}-32\pi G\gamma C_{acde}C_{b}^{~cde}+  O\left(\epsilon ^3\right).
\end{equation}
Note that since the right-hand side is non-zero only due to contributions to the equations of motion from the variation of the Gauss-Bonnet term, the Gauss-Bonnet theorem requires it to vanish for $d=4$.  It also vanishes for $d < 4$ where the Weyl tensor is identically zero.

Now consider a null hypersurface $N$ generated by a hypersurface-orthognal null normal vector field $k^a$.  For simplicity, we choose both the expansion $\theta$ and the shear $\sigma_{ab}$ of $N$ to vanish at some point $p$, or equivalently that the extrinsic curvature along $k$ vanishes there for any cut $\Sigma$ of $N$ through $p$; i.e.,
\begin{equation}
\label{eq:localstatonarity}
K^{(k)}_{ab}|_p := (\tilde h_{a}^{~c} \tilde h_{b}^{~d} \nabla_{c} k_{d})|_p=0,
\end{equation}
where $\tilde h_{a}^{~c}$ is the projector onto $\Sigma$.  As in e.g. \cite{Cao:2010vj}, we will use the notation
$K^{(X)}_{ab} := \tilde h_{a}^{~c} \tilde h_{b}^{~d} \nabla_{c} X_{d}$ below for any vector field $X_{d}$ orthogonal to $\Sigma$.
Note that \eqref{eq:localstatonarity} does not restrict the spacetime at $p$ in any way; given any $p$ in any spacetime, we may choose $\Sigma$ and then define the orthogonal null congruence $N$ so that the above conditions are satisfied. We use indices $a$, $b$, $c$, $d$, \ldots to denote coordinates in spacetime and indices $\alpha $, $\beta $, $\gamma $, $\delta $, \ldots to denote coordinates on $\Sigma$.

It is convenient to also introduce an auxiliary null vector field $l^a$ orthogonal to $\Sigma$ and satisfying $g_{ab}k^al^b=-1$.  The spacetime metric can then be written
\begin{equation}
\label{eq:metricseparation}
g_{ab}=\tilde{h}_{ab}-k_al_b-l_ak_b,
\end{equation}
where the transverse part $\tilde{h}_{ab} =\tilde{h}_{a}^{~c}g_{cb}$ is the induced metric on $\Sigma$.  We will reserve $k$ and $l$ ``indices'' to denote contractions with $k^{a}$ and $l^{a}$, as in e.g. $A_{kl} := A_{ab}k^{a}l^{b}$. Substituting \eqref{eq:metricseparation} into equation \eqref{eq:eom} and noticing that $C_{klk\alpha }=-C_{k\beta \alpha }^{~~~\beta }$ for all $d$, the Raychaudhuri equation $\dot{\theta} = - \frac{\theta^2}{d-2} - \sigma^{ab} \sigma_{ab} - R_{ab} k^a k^b$ for hypersurface-orthogonal null congruences satisfying equation \eqref{eq:localstatonarity} yields
\begin{equation}
\label{eq:Tkk}
\begin{aligned}
\dot{\theta}|_p &= -R_{ab} k^a k^b=32\pi G\gamma C_{acde}C_{b}^{~cde}k^ak^b +O\left(\epsilon ^3\right) \\
&= 32 \pi G\gamma \left(C_{k\alpha \beta \gamma }C_k^{~\alpha \beta \gamma }-2C_{k\beta \alpha }^{~~~\beta }C_{k\gamma }^{~~\alpha \gamma }-4C_{k\alpha k\beta }C_{k~l~}^{~\alpha ~\beta }\right) +O\left(\epsilon ^3\right).
\end{aligned}
\end{equation}

As noted above, \eqref{eq:Tkk} vanishes for $d=4$.  One may see this explicitly by using the $d=4$ identity $C_{k\alpha \beta \gamma }C_k^{~\alpha \beta \gamma }=2C_{k\beta \alpha }^{~~~\beta }C_{k\gamma }^{~~\alpha \gamma }$ from \cite{Coley:2009is} so that the first two terms cancel in \eqref{eq:Tkk}.  To deal with the final term we again use the $d=4$ results from \cite{Coley:2009is} to write $C_{k\alpha l\beta}$ as $C_{k\alpha l\beta}=-\frac{1}{4}A\tilde{h}_{\alpha \beta }+\frac{1}{2}B\epsilon _{\alpha \beta }$ where $\epsilon _{\alpha \beta}$ is the area element of $\Sigma$ and $A$ and $B$ are independent scalars; in particular, there is no traceless symmetric term. The final term in \eqref{eq:Tkk} then vanishes since
$C_{k\alpha k~}^{~~~\alpha }=0 = C_{k\alpha k \beta} \epsilon^{\alpha \beta}$ identically for all $d$.

To study the QFC, recall \cite{Dong:2013qoa,Camps:2013zua} that the entropy functional associated with the Gauss-Bonnet term is
\begin{equation}
\label{eq:entropy}
S_\text{GB}=-8\pi \gamma \int_\Sigma d^{d-2} y \sqrt{\tilde {h}}\tilde{R},
\end{equation}
where $\tilde{R}$ is the scalar curvature of the induced metric $\tilde{h}_{\alpha \beta}$. Let us introduce a deformation vector field $X^a=fk^a$ on $N$, where $f$ is a scalar function of the null generators $y$. Taking $f= \delta(y-y_p)$, when $\Sigma$ is is deformed along $X^a$ the first derivative of entropy \eqref{eq:entropy} is
\begin{equation}
\begin{aligned}
\label{eq:deltaXSGB}
\delta _XS_\text{GB}&=-8\pi \gamma  \int_\Sigma d^{d-2} y \sqrt{\tilde {h}}\left(\tilde{R}^{ab}-\frac{1}{2}\tilde{R}\tilde{h}^{ab}\right)\delta _X\tilde{h}_{ab}\\
&=-16\pi \gamma  \int_\Sigma d^{d-2} y\sqrt{\tilde {h}}\left(\tilde{R}^{ab}-\frac{1}{2}\tilde{R}\tilde{h}^{ab}\right)K^{(X)}_{ab}\\
&=-16\pi \gamma \sqrt{\tilde h} \left(\tilde{R}^{ab}-\frac{1}{2}\tilde{R}\tilde{h}^{ab}\right)K^{(k)}_{ab}.
\end{aligned}
\end{equation}
Here, to obtain the second line, we used $\delta _X\tilde{h}_{ab}=2K^{(X)}_{ab}$ (i.e. equation (3.10) of \cite{Cao:2010vj}).

We now introduce another vector field $Z = \delta(y-y_Z) k^a$. Recalling that $K^{(k)}_{ab}$ vanishes at $p$, we find the second derivative
\begin{equation}
\label{eq:ZXvary}
\delta _Z \left(\frac{1}{\sqrt{\tilde{h}}} \delta _XS_\text{GB}\right)=-16\pi \gamma \left(\tilde{R}^{ab}-\frac{1}{2}\tilde{R}\tilde{h}^{ab}\right)(\delta _ZK^{(k)}_{ab})|_p.
\end{equation}
Since $K^{(k)}_{ab}|_p=0$ and $Z^a=\delta(y-y_Z) k^a$, the derivative of $K^{(k)}_{ab}$ at $p$ takes the simple form \cite{Cao:2010vj}
\begin{equation}
(\delta _Z K^{(k)}_{ab})|_p  =(-\tilde{h}_a^{~c}\tilde{h}_b^{~d}Z^ek^fR_{ecfd})|_p
\end{equation}
and \eqref{eq:ZXvary} becomes $\delta _Z \left(\frac{1}{\sqrt{\tilde{h}}} \delta_X S_{\text{GB}}\right) =  \delta(y_p-y_Z)S''_{\text{GB}}$ for
\begin{equation}
\label{eq:ddS}
S_\text{GB}''=16\pi \gamma \left(\tilde{R}^{ab}-\frac{1}{2}\tilde{R}\tilde{h}^{ab}\right)R_{kakb}.
\end{equation}
Since we treat the theory classically, we save for the end of this section consideration of any explicit $S_{\rm out}$ term in equation \eqref{SgenDef} associated with the entropy of gravitons and thus find
\begin{equation}
\frac{\delta }{\delta \Sigma \left(y_Z\right)}\Theta \left[\Sigma; y_p \right] = \sqrt{\tilde{h}} Q \delta(y_p-y_Z)
\end{equation}
for
\begin{equation}
\label{Qdef}
Q = \dot{\theta} + 4GS''_{\text{GB}}.
\end{equation}

Since $K^{(k)}_{ab}|_p=0$, the Gauss equation (i.e. equation (2.14) of \cite{Cao:2010vj}) at $p$ is simply
\begin{equation}
(\tilde{R}_{abcd})|_p=(\tilde{h}_a^{~e}\tilde{h}_b^{~f}\tilde{h}_c^{~g}\tilde{h}_d^{~h}R_{efgh})|_p,
\end{equation}
and expression \eqref{eq:ddS} becomes
\begin{equation}
S_\text{GB}''=16\pi \gamma \left(R_{cedf}\tilde{h}^{cd}\tilde{h}^{ae}\tilde{h}^{bf}-\frac{1}{2}R_{cedf}\tilde{h}^{cd}\tilde{h}^{ef}\tilde{h}^{ab}\right)R_{kakb}.
\end{equation}
In the weak curvature limit, we may use \eqref{eq:weakgravity} and \eqref{eq:eom} to further write
\begin{equation}
\label{eq:S''}
\begin{aligned}
S_\text{GB}''&=16\pi \gamma \left(C_{cedf}\tilde{h}^{cd}\tilde{h}^{ae}\tilde{h}^{bf}-\frac{1}{2}C_{cedf}\tilde{h}^{cd}\tilde{h}^{ef}\tilde{h}^{ab}\right)C_{kakb}+O\left(\epsilon ^3\right)\\
&=32\pi \gamma C_{k\alpha k\beta }C_{k~l}^{~\alpha ~\beta }+O\left(\epsilon ^3\right),
\end{aligned}
\end{equation}
where in the last step we have used $\tilde{h}^{ab}C_{kakb} = C_{kkkl} + C_{klkk}$ which vanishes since the Weyl tensor is anti-symmetric in pairs of indices ($C_{abcd} = - C_{bacd} = -C_{abdc}$).  Combining \eqref{eq:Tkk} and \eqref{eq:S''} with the definition \eqref{Qdef} yields
\begin{equation}
\label{eq:qfc}
Q= 32 \pi G \gamma \left( C_{k\alpha \beta \gamma } C_k^{~\alpha \beta \gamma }-2C_{k\beta \alpha }^{~~~\beta }C_{k\gamma }^{~~\alpha \gamma }\right) +O\left(\epsilon ^3\right).
\end{equation}

As with \eqref{eq:Tkk}, expression \eqref{eq:qfc} vanishes for $d=4$. To show that it generally does not vanish for $d=5$, we use further results from \cite{Coley:2009is} to write it in terms of independent components of the Weyl tensor; the Weyl tensor at a point is constrained by its symmetries, tracelessness, and the algebraic Bianchi identity. The block $C_{k\alpha \beta \gamma }$, which has boost weight $-1$, can be written in terms of $8$ independent components as
\begin{equation}
\label{eq:block1}
C_{k\alpha \beta \gamma }=\tilde{h}_{\alpha \beta }v_\gamma -\tilde{h}_{\alpha \gamma }v_\beta +\epsilon _{\beta \gamma }^{~~\delta }n_{\delta \alpha }\text{, for }d=5,
\end{equation}
where $\epsilon_{\alpha\beta\gamma}$ is the area element of $\Sigma$, $v_\gamma $ is a vector containing $3$ independent components and $n_{\delta \alpha }$ is a traceless symmetric matrix containing $5$ independent components. Thus,
\begin{equation}
\label{eq:finalQ}
Q=64\pi G\gamma \left(n_{\alpha \beta}n^{\alpha \beta}-2v_\gamma v^\gamma \right)+  O\left(\epsilon ^3\right)\text{, for }d=5.
\end{equation}
Furthermore, for $d > 5$ we may again use \cite{Coley:2009is} to take the block $C_{k\alpha \beta \gamma}$ to be of the form \eqref{eq:block1}, although \eqref{eq:block1} is no longer the most general form for $C_{k\alpha \beta \gamma}$ and of course the number of components of each object above increases with the spacetime dimension $d$.

It is clear from \eqref{eq:finalQ} that \eqref{eq:qfc} is generally non-zero for $d\ge5$. Furthermore, while the QFC requires $Q$ to be non-positive, for $\gamma > 0$ it can be made positive by setting $v_\gamma =0$ and taking $n_{\alpha \beta} \neq 0$, and for $\gamma < 0$ we can make $Q$ positive by taking $n_{\alpha \beta} = 0$ with $v_\gamma \neq 0$.

Violations of the QFC thus occur for either sign of the Gauss-Bonnet coupling $\gamma$ and the QFC generally fails for classical $d\ge 5$ Einstein-Hilbert-Gauss-Bonnet gravity.  We may immediately extend this result to the quantum level by noting that graviton contributions to the $S_{\rm out}$ term of equation \eqref{SgenDef} are of order $G$ while our violation above is of order $G\gamma$.  The key point here is that $\gamma$ has dimensions $( \textit{Length})^{-(d-4)}$ so that the $G\gamma$ term is more important at large length scales than the $G$ term in $S_{\text{out}}$.   In other words, the classical contributions to \eqref{SgenDef} will dominate in the long-distance limit.

Let us now consider more general (perhaps, effective) theories of gravity with higher derivative terms. First, it is trivial to add a cosmological constant $\Lambda$ to the action \eqref{eq:action}.  Noticing that $C_{k\alpha k}^{~~~\alpha }=0$ identically for all $d$, one finds no change to equation \eqref{eq:S''}. Next, recall that at the four-derivative level, up to total derivatives there are only two further independent terms that we may add to the action, and we may choose to write both in terms of the square of the Ricci tensor (so that they do not depend on the Weyl tensor).   Thus Ricci-flat metrics continue to solve the theory with $\gamma =0$, and there continue to be solutions of the form \eqref{eq:eom} in the presence of such terms, and in such cases we again find \eqref{eq:qfc} (up to additional corrections that are also of order $\epsilon^2$ but involve additional derivatives and so remains smaller in the long-distance limit).  Finally, so long as they are controlled by a common length scale, in a long-distance expansion any terms in the action with more than four derivatives can be ignored relative to those already discussed so that \eqref{eq:eom} continues to hold in that regime.

The key point, however, is the associated implication for generic quantum theories of massive fields when coupled to semi-classical gravity.   Since integrating out massive fields gives an effective action of the above type, so long as the resulting Gauss-Bonnet coefficient\footnote{The final Gauss-Bonnet coefficient is of course formally the sum of the Gauss-Bonnet coefficient in the gravitational action and the coefficient induced by integrating out the matter.  For $d \ge 5$ the latter is generally divergent, so the former must be as well if the effective action is to be finite.  In this sense, as usual, there is generally no meaning to attempting to couple the massive field theory to Einstein-Hilbert gravity alone.} is non-zero the theory will violate the associated QFC.

\section{Violating the Generalized Covariant Entropy Bound}
\label{Bousso}

Bousso's original covariant entropy bound \cite{Bousso:1999xy}  involved the concept of ``entropy flux through a non-expanding null surface" and conjectured this to be bounded by $(\frac{1}{4G}$ times) the area of the largest cut. There has been much discussion of how this concept might be properly defined, with one seemingly-natural choice involving entropy defined directly on the null surface. This version was proven for free and interacting theories in the $G \to 0$ limit from the monotonicity property of the relative entropy \cite{Bousso:2014sda,Bousso:2014uxa}.  Alternatively, Strominger and Thompson \cite{Strominger:2003br} suggested focussing on the case where any cut of the null surface $N$ is closed and bounds a spacelike surface.  One may then discuss the von Neumann entropy $S_{vN}$ of the region enclosed, and replace the ``flux of entropy across $N$" with the change in $S_{vN}$ between the initial and final surfaces.

As noted in \cite{Bousso:2015mna}, this choice gives rise to a putative (generalized) covariant entropy bound which is intrinsically finite and does not require renormalization.  The conjecture of \cite{Bousso:2015mna} states that if some set of null generators has non-positive {\it quantum} expansion ($\Theta \le 0$) on some cut $C_{\text{initial}}$ of $N$, then any cut $C_{\text{final}}$ obtained by moving $C_{\text{initial}}$ to the future along these generators will have smaller or equal generalized entropy $S_{\text{gen}}$ so long as no caustic lies between $C_{\text{initial}}$ and $C_{\text{final}}$.  The non-increase of $S_{\text{gen}}$ is equivalent to the claim
\begin{equation}
\label{eq:GenCEB}
\Delta S \le \Delta A/4G,
\end{equation}
which is a generalized covariant entropy bound of the form first discussed in \cite{Flanagan:1999jp}.  Note, however, that the condition $\Theta|_{C_{\text{initial}}} \le 0$   under which this was conjectured in \cite{Bousso:2015mna} differs from the assumption used in \cite{Bousso:1999xy,Flanagan:1999jp} which requires the classical expansion $\theta$ to be non-positive on all intermediate cuts.  Furthermore, equation \eqref{eq:GenCEB} follows directly from the QFC in cases where the latter is valid \cite{Bousso:2015mna}.

However, it turns out the QFC violation constructed above is also a counterexample to the generalized covariant entropy bound (i.e. the quantum Bousso bound) of \cite{Bousso:2015mna}.  The key point is that the Gauss-Bonnet contribution \eqref{eq:deltaXSGB} to the the quantum expansion vanishes at $p$ since $K_{ab}^{(X)} =0$.  But since $\theta|_p=0$ as well, the full quantum expansion $\Theta$ also vanishes at $p$.

From here we need only note that we can then achieve $\Theta \le 0$ near $p$ on $C_{\text{initial}}$ by taking the classical expansion $\theta$ sufficiently negative near $p$; i.e., by simply choosing $C_{\text{initial}}$ to have large enough extrinsic curvature of the appropriate sign.  We then find that later cuts $C_{\text{final}}$ differing from $C_{\text{initial}}$ only very near $p$ and by small affine parameter distance along the QFC-violating generators must have larger generalized entropy $S_{\text{gen}}$, violating the conjecture of \cite{Bousso:2015mna}.  Indeed, in the appropriate limit the increase of $S_{\text{gen}}$ is determined by $\left( \pounds_k \Theta \right) |_p > 0$.

\section{Discussion}
\label{sec:Discussion}

Using an explicit calculation for classical Einstein-Hilbert-Gauss-Bonnet gravity, we argued that the QFC of \cite{Bousso:2015mna} is violated in generic $d \ge 5$ theories of gravity coupled to massive quantum fields. The key point is that integrating out the massive fields generically induces a Gauss-Bonnet term which, at least for a certain class of solutions, dominates in the long-distance limit.  There we may use the explicit Einstein-Hilbert-Gauss-Bonnet calculation of section \ref{sec:Violation}.  We expect similar violations to continue to arise when massless quantum fields are included as well.  Our construction also provides a counterexample to the generalized covariant entropy bound (i.e. the quantum Bousso bound) conjectured in \cite{Bousso:2015mna}. It remains an open question whether the QFC and covariant entropy bound could hold for $d \le 4$, and it would be interesting to investigate the affect of Ricci-squared terms in this context.
As mentioned in the introduction, the QFC is closely related to the Quantum Null Energy Condition (QNEC).  Indeed, when a matter theory satisfying the Quantum Null Energy Condition is coupled to Einstein-Hilbert gravity, the QFC will hold at least to first order in the gravitational coupling $G$.  The reader may thus ask whether our results are in tension with the QNEC proofs in \cite{Bousso:2015wca} and \cite{Koeller:2015qmn}.  The answer is no, as those results prove the QNEC only for congruences $N$ through $p$ that form bifurcate Killing horizons at $G=0$.  And on a bifurcate Killing horizon components of the Weyl tensor with non-zero boost weight must vanish.  This would then force $C_{k\alpha \beta \gamma}=0$ and thus $Q=0$ in \eqref{eq:qfc}, reproducing the expected result that the QFC hold at first order in $G\gamma $ for such cases.\footnote{Indeed, a result of \cite{Kolekar:2012tq} shows that the QFC holds for any Lovelock theory of gravity (a class which includes the Einstein-Hilbert-Gauss-Bonnet gravity) when evaluated at first order in $G$ about a Killing horizon.  This result was then generalized in \cite{Sarkar:2013swa} and extended to arbitrary higher-derivative theories of gravity in \cite{Wall:2015raa}.}

Conversely, taking the limit $G\gamma \rightarrow 0$ of our results shows that for $d \ge 5$ the renormalized QNEC must generally fail\footnote{As will be discussed in more detail in \cite{ToAppear}, the QNEC may still hold in some sense for appropriate bare quantities.  But finite renormalized quantities cannot satisfy a QNEC-like bound.} for surfaces $\Sigma$ defining null congruences $N$ that are only locally stationary at $p$; i.e., which satisfy $\theta = \sigma_{ab} = R_{ab} k^a k^b =0$ in the background spacetime. However, one may ask if the QNEC can hold at locally stationary points of null congruences for $d < 5$ or where further conditions are satisfied. The forthcoming work \cite{ToAppear} will provide results of this kind, including a proof for $d \le 3$ holographic theories at locally stationary points.

It is natural to ask if our QFC violation also provides a perturbative counterexample to the GSL.  While Einstein-Hilbert-Gauss-Bonnet gravity is known to violate the GSL at the non-perturbative level \cite{Jacobson:1993xs,Liko:2007vi,Sarkar:2010xp}, these are of lesser interest as higher derivative theories of gravity are expected \cite{Camanho:2014apa} to approximate UV-complete theories only when treated perturbatively as an effective field theory valid at lengths longer than some cutoff scale $\ell_c$. And indeed, as in section \ref{Bousso}, one can certainly find cases where the generalized entropy inside the horizon increases and thus that outside decreases.  But the GSL is naturally conjectured to hold at most for causal horizons (see e.g. \cite{Jacobson:2003wv}, \cite{Wall:2011hj}), and determining whether a given null $N$ is a causal horizon requires understanding the very far future.  Analyzing the constraints on $N$, thus requires going well beyond the local approximations used here, and thus beyond the scope of this work, though see \cite{Wall:2015raa,Bhattacharjee:2015qaa,Bhattacharyya:2016xfs} for further work on the GSL for higher derivative gravity and more thorough reviews.

Finally, one may ask if some version of the QFC or quantum Bousso bound might yet be salvaged for general $d \ge 5$ theories.  In particular, we recall again that higher derivative gravity should be treated as an effective field theory with a cutoff $\ell_c$.  But the QFC, and in particular our construction of a counterexample, requires the choice of a null congruence $N$ that is taken to be arbitrarily well localized in the transverse directions.  Furthermore, since the Gauss-Bonnet term should be treated as perturbatively small, correspondingly small changes in $N$ can make $\theta, \sigma_{ab}$ non zero at $p$ so that $\dot{\theta} = - \frac{\theta^2}{d-2} - \sigma^{ab} \sigma_{ab} - R_{ab} k^a k^b$ becomes sufficiently negative at $p$ that $Q < 0$ for the new surface.  In other words, perturbatively close to any compact QFC-violating null congruence $N$ lies a QFC-respecting null congruence $N'$.  If this can be interpreted as a distinction finer than the cutoff scale $\ell_c$, there is room for the formulation of an effective QFC valid only at larger scales.\footnote{We thank Aron Wall for this suggestion.} But such an interpretation is not immediately clear as the above mentioned deformation from $N$ to $N'$ involves adding extrinsic curvature of a particular sign; it is not just a transverse smearing of the surface.  And while it is attractive from many perspectives to conjecture that a QFC-like inequality may hold in an appropriately cutoff sense, both the form that this effective QFC might take and how in practice it would be used to restrict possible pathologies of NEC-violating spacetimes remain open questions for future investigation.

\paragraph{Note added in v2.}After the appearance of our paper on the arXiv, it was pointed out in \cite{Leichenauer:2017bmc} that the violation described above is removed by restricting the QFC to apply only to variations of the entropy defined by surfaces that are smooth on the scale set by $G\gamma$, and which is presumably associated with the cut-off that defines the effective theory.  This emphasizes the importance of studying the effect of $R_{ab}R^{ab}$ terms in the action, which might contribute a different class of terms to the QFC.

\section*{Acknowledgements}
It is a pleasure to thank Chris Akers, Raphael Bousso, Venkatesh Chandrasekaran, Netta Engelhardt, Zachary Fisher, Stefan Leichenauer, Adam Levine, Arvin Shahbazi Moghaddam, and Aron Wall for useful discussions. ZF and DM were supported in part by the Simons Foundation and by funds from the University of California. JK was supported in part by the Berkeley Center for Theoretical Physics, by the National Science Foundation (award numbers 1521446, and 1316783), by FQXi, and by the US Department of Energy under contract DE-AC02-05CH11231.

\bibliographystyle{jhep}
	\cleardoublepage
\phantomsection
\renewcommand*{\bibname}{References}

\bibliography{references}

\providecommand{\href}[2]{#2}\begingroup\raggedright\begin{thebibliography}{10}

\bibitem{Wald:1984rg}
R.~M. Wald, \emph{{General Relativity}}.
\newblock University of Chicago Press, 1984.

\bibitem{Hawking:1971tu}
S.~W. Hawking, \emph{{Gravitational Radiation from Colliding Black Holes}},
  \href{http://dx.doi.org/10.1103/PhysRevLett.26.1344}{\emph{Phys. Rev. Lett.}
  {\bfseries 26} (1971) 1344--1346}.

\bibitem{Penrose:1964wq}
R.~Penrose, \emph{{Gravitational Collapse and Space-Time Singularities}},
  \href{http://dx.doi.org/10.1103/PhysRevLett.14.57}{\emph{Phys. Rev. Lett.}
  {\bfseries 14} (1965) 57--59}.

\bibitem{Hawking:1973uf}
S.~W. Hawking and G.~F.~R. Ellis, \emph{{The Large Scale Structure of
  Space-Time}}.
\newblock Cambridge University Press, 1973.

\bibitem{Hawking:1991nk}
S.~W. Hawking, \emph{{The Chronology Protection Conjecture}},
  \href{http://dx.doi.org/10.1103/PhysRevD.46.603}{\emph{Phys. Rev.} {\bfseries
  D46} (1992) 603--611}.

\bibitem{Friedman:1993ty}
J.~L. Friedman, K.~Schleich and D.~M. Witt, \emph{{Topological Censorship}},
  \href{http://dx.doi.org/10.1103/PhysRevLett.71.1486}{\emph{Phys. Rev. Lett.}
  {\bfseries 71} (1993) 1486--1489},
  [\href{https://arxiv.org/abs/gr-qc/9305017}{{\ttfamily gr-qc/9305017}}].

\bibitem{Wall:2012uf}
A.~C. Wall, \emph{{Maximin Surfaces, and the Strong Subadditivity of the
  Covariant Holographic Entanglement Entropy}},
  \href{http://dx.doi.org/10.1088/0264-9381/31/22/225007}{\emph{Class. Quant.
  Grav.} {\bfseries 31} (2014) 225007},
  [\href{https://arxiv.org/abs/1211.3494}{{\ttfamily 1211.3494}}].

\bibitem{Headrick:2014cta}
M.~Headrick, V.~E. Hubeny, A.~Lawrence and M.~Rangamani, \emph{{Causality {\&}
  Holographic Entanglement Entropy}},
  \href{http://dx.doi.org/10.1007/JHEP12(2014)162}{\emph{JHEP} {\bfseries 12}
  (2014) 162}, [\href{https://arxiv.org/abs/1408.6300}{{\ttfamily 1408.6300}}].

\bibitem{Epstein:1965zza}
H.~Epstein, V.~Glaser and A.~Jaffe, \emph{{Nonpositivity of Energy Density in
  Quantized Field Theories}},
  \href{http://dx.doi.org/10.1007/BF02749799}{\emph{Nuovo Cim.} {\bfseries 36}
  (1965) 1016}.

\bibitem{Gao:2016bin}
P.~Gao, D.~L. Jafferis and A.~Wall, \emph{{Traversable Wormholes via a Double
  Trace Deformation}},  \href{https://arxiv.org/abs/1608.05687}{{\ttfamily
  1608.05687}}.

\bibitem{Wall:2013uza}
A.~C. Wall, \emph{{The Generalized Second Law implies a Quantum Singularity
  Theorem}}, \href{http://dx.doi.org/10.1088/0264-9381/30/19/199501,
  10.1088/0264-9381/30/16/165003}{\emph{Class. Quant. Grav.} {\bfseries 30}
  (2013) 165003}, [\href{https://arxiv.org/abs/1010.5513}{{\ttfamily
  1010.5513}}].

\bibitem{Bousso:1999xy}
R.~Bousso, \emph{{A Covariant entropy conjecture}},
  \href{http://dx.doi.org/10.1088/1126-6708/1999/07/004}{\emph{JHEP} {\bfseries
  07} (1999) 004}, [\href{https://arxiv.org/abs/hep-th/9905177}{{\ttfamily
  hep-th/9905177}}].

\bibitem{Bousso:2015mna}
R.~Bousso, Z.~Fisher, S.~Leichenauer and A.~C. Wall, \emph{{Quantum focusing
  conjecture}}, \href{http://dx.doi.org/10.1103/PhysRevD.93.064044}{\emph{Phys.
  Rev.} {\bfseries D93} (2016) 064044},
  [\href{https://arxiv.org/abs/1506.02669}{{\ttfamily 1506.02669}}].

\bibitem{Strominger:2003br}
A.~Strominger and D.~M. Thompson, \emph{{A Quantum Bousso bound}},
  \href{http://dx.doi.org/10.1103/PhysRevD.70.044007}{\emph{Phys. Rev.}
  {\bfseries D70} (2004) 044007},
  [\href{https://arxiv.org/abs/hep-th/0303067}{{\ttfamily hep-th/0303067}}].

\bibitem{Dong:2013qoa}
X.~Dong, \emph{{Holographic Entanglement Entropy for General Higher Derivative
  Gravity}}, \href{http://dx.doi.org/10.1007/JHEP01(2014)044}{\emph{JHEP}
  {\bfseries 01} (2014) 044},
  [\href{https://arxiv.org/abs/1310.5713}{{\ttfamily 1310.5713}}].

\bibitem{Miao:2014nxa}
R.-X. Miao and W.-z. Guo, \emph{{Holographic Entanglement Entropy for the Most
  General Higher Derivative Gravity}},
  \href{http://dx.doi.org/10.1007/JHEP08(2015)031}{\emph{JHEP} {\bfseries 08}
  (2015) 031}, [\href{https://arxiv.org/abs/1411.5579}{{\ttfamily 1411.5579}}].

\bibitem{Dong:2015zba}
X.~Dong and R.-X. Miao, \emph{{Generalized Gravitational Entropy from Total
  Derivative Action}},
  \href{http://dx.doi.org/10.1007/JHEP12(2015)100}{\emph{JHEP} {\bfseries 12}
  (2015) 100}, [\href{https://arxiv.org/abs/1510.04273}{{\ttfamily
  1510.04273}}].

\bibitem{Wall:2015raa}
A.~C. Wall, \emph{{A Second Law for Higher Curvature Gravity}},
  \href{http://dx.doi.org/10.1142/S0218271815440149}{\emph{Int. J. Mod. Phys.}
  {\bfseries D24} (2015) 1544014},
  [\href{https://arxiv.org/abs/1504.08040}{{\ttfamily 1504.08040}}].

\bibitem{Jacobson:1993xs}
T.~Jacobson and R.~C. Myers, \emph{{Black hole entropy and higher curvature
  interactions}},
  \href{http://dx.doi.org/10.1103/PhysRevLett.70.3684}{\emph{Phys. Rev. Lett.}
  {\bfseries 70} (1993) 3684--3687},
  [\href{https://arxiv.org/abs/hep-th/9305016}{{\ttfamily hep-th/9305016}}].

\bibitem{will_lattice}
W.~Donnelly, \emph{{Decomposition of entanglement entropy in lattice gauge
  theory}}, \href{http://dx.doi.org/10.1103/PhysRevD.85.085004}{\emph{Phys.
  Rev.} {\bfseries D85} (2012) 085004},
  [\href{https://arxiv.org/abs/1109.0036}{{\ttfamily 1109.0036}}].

\bibitem{will14}
W.~Donnelly, \emph{{Entanglement entropy and nonabelian gauge symmetry}},
  \href{http://dx.doi.org/10.1088/0264-9381/31/21/214003}{\emph{Class. Quant.
  Grav.} {\bfseries 31} (2014) 214003},
  [\href{https://arxiv.org/abs/1406.7304}{{\ttfamily 1406.7304}}].

\bibitem{Ghosh:2015iwa}
S.~Ghosh, R.~M. Soni and S.~P. Trivedi, \emph{{On The Entanglement Entropy For
  Gauge Theories}},
  \href{http://dx.doi.org/10.1007/JHEP09(2015)069}{\emph{JHEP} {\bfseries 09}
  (2015) 069}, [\href{https://arxiv.org/abs/1501.02593}{{\ttfamily
  1501.02593}}].

\bibitem{Aoki:2015bsa}
S.~Aoki, T.~Iritani, M.~Nozaki, T.~Numasawa, N.~Shiba and H.~Tasaki, \emph{{On
  the definition of entanglement entropy in lattice gauge theories}},
  \href{http://dx.doi.org/10.1007/JHEP06(2015)187}{\emph{JHEP} {\bfseries 06}
  (2015) 187}, [\href{https://arxiv.org/abs/1502.04267}{{\ttfamily
  1502.04267}}].

\bibitem{Donnelly:2014fua}
W.~Donnelly and A.~C. Wall, \emph{{Entanglement entropy of electromagnetic edge
  modes}}, \href{http://dx.doi.org/10.1103/PhysRevLett.114.111603}{\emph{Phys.
  Rev. Lett.} {\bfseries 114} (2015) 111603},
  [\href{https://arxiv.org/abs/1412.1895}{{\ttfamily 1412.1895}}].

\bibitem{Donnelly:2015hxa}
W.~Donnelly and A.~C. Wall, \emph{{Geometric entropy and edge modes of the
  electromagnetic field}},
  \href{http://dx.doi.org/10.1103/PhysRevD.94.104053}{\emph{Phys. Rev.}
  {\bfseries D94} (2016) 104053},
  [\href{https://arxiv.org/abs/1506.05792}{{\ttfamily 1506.05792}}].

\bibitem{Bousso:2015wca}
R.~Bousso, Z.~Fisher, J.~Koeller, S.~Leichenauer and A.~C. Wall, \emph{{Proof
  of the Quantum Null Energy Condition}},
  \href{http://dx.doi.org/10.1103/PhysRevD.93.024017}{\emph{Phys. Rev.}
  {\bfseries D93} (2016) 024017},
  [\href{https://arxiv.org/abs/1509.02542}{{\ttfamily 1509.02542}}].

\bibitem{Koeller:2015qmn}
J.~Koeller and S.~Leichenauer, \emph{{Holographic Proof of the Quantum Null
  Energy Condition}},
  \href{http://dx.doi.org/10.1103/PhysRevD.94.024026}{\emph{Phys. Rev.}
  {\bfseries D94} (2016) 024026},
  [\href{https://arxiv.org/abs/1512.06109}{{\ttfamily 1512.06109}}].

\bibitem{Shapiro:2008sf}
I.~L. Shapiro, \emph{{Effective Action of Vacuum: Semiclassical Approach}},
  \href{http://dx.doi.org/10.1088/0264-9381/25/10/103001}{\emph{Class. Quant.
  Grav.} {\bfseries 25} (2008) 103001},
  [\href{https://arxiv.org/abs/0801.0216}{{\ttfamily 0801.0216}}].

\bibitem{Camanho:2014apa}
X.~O. Camanho, J.~D. Edelstein, J.~Maldacena and A.~Zhiboedov, \emph{{Causality
  Constraints on Corrections to the Graviton Three-Point Coupling}},
  \href{http://dx.doi.org/10.1007/JHEP02(2016)020}{\emph{JHEP} {\bfseries 02}
  (2016) 020}, [\href{https://arxiv.org/abs/1407.5597}{{\ttfamily 1407.5597}}].

\bibitem{Cao:2010vj}
L.-M. Cao, \emph{{Deformation of Codimension-2 Surface and Horizon
  Thermodynamics}},
  \href{http://dx.doi.org/10.1007/JHEP03(2011)112}{\emph{JHEP} {\bfseries 03}
  (2011) 112}, [\href{https://arxiv.org/abs/1009.4540}{{\ttfamily 1009.4540}}].

\bibitem{Coley:2009is}
A.~Coley and S.~Hervik, \emph{{Higher dimensional bivectors and classification
  of the Weyl operator}},
  \href{http://dx.doi.org/10.1088/0264-9381/27/1/015002}{\emph{Class. Quant.
  Grav.} {\bfseries 27} (2010) 015002},
  [\href{https://arxiv.org/abs/0909.1160}{{\ttfamily 0909.1160}}].

\bibitem{Camps:2013zua}
J.~Camps, \emph{{Generalized Entropy and Higher Derivative Gravity}},
  \href{http://dx.doi.org/10.1007/JHEP03(2014)070}{\emph{JHEP} {\bfseries 03}
  (2014) 070}, [\href{https://arxiv.org/abs/1310.6659}{{\ttfamily 1310.6659}}].

\bibitem{Bousso:2014sda}
R.~Bousso, H.~Casini, Z.~Fisher and J.~Maldacena, \emph{{Proof of a Quantum
  Bousso Bound}},
  \href{http://dx.doi.org/10.1103/PhysRevD.90.044002}{\emph{Phys. Rev.}
  {\bfseries D90} (2014) 044002},
  [\href{https://arxiv.org/abs/1404.5635}{{\ttfamily 1404.5635}}].

\bibitem{Bousso:2014uxa}
R.~Bousso, H.~Casini, Z.~Fisher and J.~Maldacena, \emph{{Entropy on a null
  surface for interacting quantum field theories and the Bousso bound}},
  \href{http://dx.doi.org/10.1103/PhysRevD.91.084030}{\emph{Phys. Rev.}
  {\bfseries D91} (2015) 084030},
  [\href{https://arxiv.org/abs/1406.4545}{{\ttfamily 1406.4545}}].

\bibitem{Flanagan:1999jp}
E.~E. Flanagan, D.~Marolf and R.~M. Wald, \emph{{Proof of classical versions of
  the Bousso entropy bound and of the generalized second law}},
  \href{http://dx.doi.org/10.1103/PhysRevD.62.084035}{\emph{Phys. Rev.}
  {\bfseries D62} (2000) 084035},
  [\href{https://arxiv.org/abs/hep-th/9908070}{{\ttfamily hep-th/9908070}}].

\bibitem{Kolekar:2012tq}
S.~Kolekar, T.~Padmanabhan and S.~Sarkar, \emph{{Entropy Increase during
  Physical Processes for Black Holes in Lanczos-Lovelock Gravity}},
  \href{http://dx.doi.org/10.1103/PhysRevD.86.021501}{\emph{Phys. Rev.}
  {\bfseries D86} (2012) 021501},
  [\href{https://arxiv.org/abs/1201.2947}{{\ttfamily 1201.2947}}].

\bibitem{Sarkar:2013swa}
S.~Sarkar and A.~C. Wall, \emph{{Generalized second law at linear order for
  actions that are functions of Lovelock densities}},
  \href{http://dx.doi.org/10.1103/PhysRevD.88.044017}{\emph{Phys. Rev.}
  {\bfseries D88} (2013) 044017},
  [\href{https://arxiv.org/abs/1306.1623}{{\ttfamily 1306.1623}}].

\bibitem{ToAppear}
Z.~Fu, J.~Koeller and D.~Marolf, ``The quantum null energy condition in curved
  space.'' 2017.

\bibitem{Liko:2007vi}
T.~Liko, \emph{{Topological deformation of isolated horizons}},
  \href{http://dx.doi.org/10.1103/PhysRevD.77.064004}{\emph{Phys. Rev.}
  {\bfseries D77} (2008) 064004},
  [\href{https://arxiv.org/abs/0705.1518}{{\ttfamily 0705.1518}}].

\bibitem{Sarkar:2010xp}
S.~Sarkar and A.~C. Wall, \emph{{Second Law Violations in Lovelock Gravity for
  Black Hole Mergers}},
  \href{http://dx.doi.org/10.1103/PhysRevD.83.124048}{\emph{Phys. Rev.}
  {\bfseries D83} (2011) 124048},
  [\href{https://arxiv.org/abs/1011.4988}{{\ttfamily 1011.4988}}].

\bibitem{Jacobson:2003wv}
T.~Jacobson and R.~Parentani, \emph{{Horizon entropy}},
  \href{http://dx.doi.org/10.1023/A:1023785123428}{\emph{Found. Phys.}
  {\bfseries 33} (2003) 323--348},
  [\href{https://arxiv.org/abs/gr-qc/0302099}{{\ttfamily gr-qc/0302099}}].

\bibitem{Wall:2011hj}
A.~C. Wall, \emph{{A proof of the generalized second law for rapidly changing
  fields and arbitrary horizon slices}},
  \href{http://dx.doi.org/10.1103/PhysRevD.87.069904,
  10.1103/PhysRevD.85.104049}{\emph{Phys. Rev.} {\bfseries D85} (2012) 104049},
  [\href{https://arxiv.org/abs/1105.3445}{{\ttfamily 1105.3445}}].

\bibitem{Bhattacharjee:2015qaa}
S.~Bhattacharjee, A.~Bhattacharyya, S.~Sarkar and A.~Sinha, \emph{{Entropy
  functionals and c-theorems from the second law}},
  \href{http://dx.doi.org/10.1103/PhysRevD.93.104045}{\emph{Phys. Rev.}
  {\bfseries D93} (2016) 104045},
  [\href{https://arxiv.org/abs/1508.01658}{{\ttfamily 1508.01658}}].

\bibitem{Bhattacharyya:2016xfs}
S.~Bhattacharyya, F.~M. Haehl, N.~Kundu, R.~Loganayagam and M.~Rangamani,
  \emph{{Towards a second law for Lovelock theories}},
  \href{http://dx.doi.org/10.1007/JHEP03(2017)065}{\emph{JHEP} {\bfseries 03}
  (2017) 065}, [\href{https://arxiv.org/abs/1612.04024}{{\ttfamily
  1612.04024}}].

\bibitem{Leichenauer:2017bmc}
S.~Leichenauer, \emph{{The Quantum Focusing Conjecture Has Not Been Violated}},
   \href{https://arxiv.org/abs/1705.05469}{{\ttfamily 1705.05469}}.

\end{thebibliography}\endgroup

\end{document}